# Information leakage resistant quantum dialogue with single photons in both polarization and spatial-mode degrees of freedom

Tian-Yu Ye*, Hong-Kun Li, Jia-Li Hu

College of Information & Electronic Engineering, Zhejiang Gongshang University, Hangzhou 310018, P.R.China

**Abstract:** In this paper, a novel quantum dialogue (QD) protocol is proposed based on single photons in both polarization and spatial-mode degrees of freedom. In the proposed QD protocol, the initial states of single photons in both polarization and spatial-mode degrees of freedom used for encoding are privately shared between two communicants through the direct transmissions of their auxiliary counterparts from one communicant to another. As a result, the information leakage problem is avoided. Moreover, the detailed security analysis also shows that the proposed QD protocol can resist Eve's several famous active attacks, such as the Trojan horse attack, the intercept-resend attack, the measure-resend attack and the entangle-measure attack. The proposed QD protocol only needs single photons in both polarization and spatial-mode degrees of freedom as quantum resource and adopts single-photon measurements. As a result, it is feasible in practice as the preparation and the measurement of a single photon in both polarization and spatial-mode degrees of freedom can be accomplished with current experimental techniques.
**Keywords:** Quantum dialogue (QD); information leakage; single photon; polarization degree of freedom; spatial-mode degree of freedom

## 1 Introduction

In the realm of quantum cryptography, quantum key distribution (QKD) [1-4] aims to create a shared random key between two communicants. Different from QKD, quantum secure direct communication (QSDC) [5-14], which is another important branch of quantum cryptography, can accomplish the direct transmission of a private message from one communicant to another without establishing a prior key to encrypt and decrypt it beforehand. However, QSDC cannot accomplish the mutual exchange of private messages between two communicants. Fortunately, in 2004, Zhang et al. [15-16] and Nguyen [17] successfully solved the above problem by independently putting forward the novel concept of quantum dialogue (QD). In other words, in a QD protocol, two communicants can easily exchange their respective messages. Subsequently, numerous QD protocols [18-24] were designed with different quantum technologies and quantum states.

However, Tan and Cai [25] and Gao et al.[26-27] independently pointed out that the phenomenon of classical correlation or information leakage exists in QD, which makes the outside Eve easily learn partial information about communicants' private messages just from the public announcement. Hereafter, researchers were devoted to designing the QD protocol without information leakage. In 2009, Shi et al. [28] suggested an information leakge resistant QD protocol through the direct transmission of a shared private Bell state. In 2010, Shi et al. put forward an information leakge resistant QD protocol through the direct transmission of a shared private single photon [29] and an information leakge resistant QD protocol based on the correlation extractability of Bell state and the auxiliary single particle [30]; Gao [31] put forward two information leakge resistant QD protocols based on the measurement correlation from the entanglement swapping between two Bell states. In 2013, the authors constructed a large payload QD protocol without information leakage based on the entanglement swapping between any two GHZ states and the auxiliary GHZ state [32]. In 2014, the authors designed an information leakge resistant QD protocol with quantum encryption [33]. In 2015, the authors put forward a kind of QD protocols without information leakage assisted by auxiliary quantum operation [34].

Recently, different from those working with the polarization states of photons, several quantum cryptography protocols working with single photons in both polarization and spatial-mode degrees of freedom [14,35-37] were proposed. It is naturally that in a quantum communication protocol, the capacity of quantum communication may be improved if the single photons in one degree of freedom are substituted by those in two degrees of freedom.

Based on the above analysis, in this paper, we are devoted to designing a novel information leakge resistant QD protocol with single photons in both polarization and spatial-mode degrees of freedom. The rest of this paper is arranged as follows: the proposed QD protocol is described in Sect.2; its security is validated in Sect.3; and finally, discussion and conclusion are given in Sect.4.

## 2 Description of QD protocols

A single-photon state in both polarization and spatial-mode degrees of freedom can be depicted as [14]

*Corresponding author:
 E-mail：happyyty@aliyun.com

$$|\phi\rangle = |\phi\rangle_P \otimes |\phi\rangle_S, \qquad (1)$$

where $|\phi\rangle_P$ and $|\phi\rangle_S$ are the single-photon states in the polarization and the spatial-mode degrees of freedom, respectively. $Z_P = \{|H\rangle, |V\rangle\}$ and $X_P = \{|R\rangle, |A\rangle\}$ are the two nonorthogonal measuring bases in the polarization degree of freedom, respectively. Here,

$$|R\rangle = \frac{1}{\sqrt{2}}(|H\rangle + |V\rangle), \quad |A\rangle = \frac{1}{\sqrt{2}}(|H\rangle - |V\rangle), \qquad (2)$$

where $|H\rangle$ and $|V\rangle$ denote the horizontal and the vertical polarizations of photons, respectively. $Z_S = \{|b_1\rangle, |b_2\rangle\}$ and $X_S = \{|s\rangle, |a\rangle\}$ are the two nonorthogonal measuring bases in the spatial-mode degree of freedom, respectively. Here,

$$|s\rangle = \frac{1}{\sqrt{2}}(|b_1\rangle + |b_2\rangle), \quad |a\rangle = \frac{1}{\sqrt{2}}(|b_1\rangle - |b_2\rangle), \qquad (3)$$

where $|b_1\rangle$ and $|b_2\rangle$ denote the upper and the lower spatial modes of photons, respectively.

The single-photon state $|\phi\rangle = |\phi\rangle_P \otimes |\phi\rangle_S$ can be produced with a 50:50 beam splitter (BS) in principle. Concretely speaking, a sequence of single-photon polarization state $|\phi\rangle_P$ is first prepared, and the spatial-mode states $|\phi\rangle_S$ are generated by BS, which is shown in Fig.1. What the BS does is to finish the transformations of the single-photon states in the spatial-mode degree of freedom [14]. The quantum states in the polarization degree of freedom and those in the spatial-mode degree of freedom are commutative, thus they can be operated independently [14].

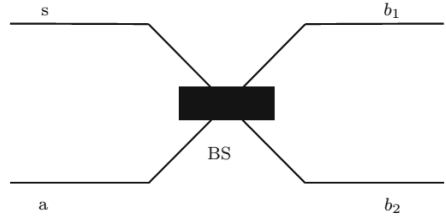

Fig.1  Schematic diagram of a Hadamard operation on a spatial quantum state of a single photon with beam splitter

Two interesting unitary operations in the polarization degree of freedom are

$$I_P = |H\rangle\langle H| + |V\rangle\langle V|, U_P = |V\rangle\langle H| - |H\rangle\langle V|, \qquad (4)$$

which cannot change the base of photon, as

$$I_P|H\rangle = |H\rangle, I_P|V\rangle = |V\rangle, I_P|R\rangle = |R\rangle, I_P|A\rangle = |A\rangle, \qquad (5)$$

$$U_P|H\rangle = |V\rangle, U_P|V\rangle = -|H\rangle, U_P|R\rangle = -|A\rangle, U_P|A\rangle = |R\rangle. \qquad (6)$$

Similarly, two interesting unitary operations in the spatial-mode degree of freedom are

$$I_S = |b_1\rangle\langle b_1| + |b_2\rangle\langle b_2|, U_S = |b_2\rangle\langle b_1| - |b_1\rangle\langle b_2|, \qquad (7)$$

which cannot change the base of photon either, as

$$I_S|b_1\rangle = |b_1\rangle, I_S|b_2\rangle = |b_2\rangle, I_S|s\rangle = |s\rangle, I_S|a\rangle = |a\rangle, \qquad (8)$$

$$U_S|b_1\rangle = |b_2\rangle, U_S|b_2\rangle = -|b_1\rangle, U_S|s\rangle = -|a\rangle, U_S|a\rangle = |s\rangle. \qquad (9)$$

Suppose that Alice has a secret consisting of $2N$ classical bits, i.e.,

$$\{(i_1, j_1), (i_2, j_2), \cdots, (i_n, j_n), \cdots, (i_N, j_N)\}, \qquad (10)$$

and Bob also has a secret consisting of $2N$ classcial bits, i.e.,

$$\{(k_1, l_1), (k_2, l_2), \cdots, (k_n, l_n), \cdots, (k_N, l_N)\}, \qquad (11)$$

where $i_n, j_n, k_n, l_n \in \{0,1\}$, $n \in \{1, 2, \cdots, N\}$. They agree on beforehand that each of the following four composite unitary operations corresponds to two classical bits such as

$$C_{00} = I_P \otimes I_S \to 00, C_{01} = I_P \otimes U_S \to 01, C_{10} = U_P \otimes I_S \to 10, C_{11} = U_P \otimes U_S \to 11. \qquad (12)$$

Inspired by Shi et al.'s QD protocol [29] and Wang et al.'s QD protocol [35], we put forward an information leakage resistant QD protocol with single photons in both polarization and spatial-mode degrees of freedom as follows.

**Step 1: Bob's preparation and transmission.** Bob prepares a sequence of $2N$ single photons in both polarization and spatial-mode degrees of freedom, i.e.,

$$S = \{L_1, L_1', L_2, L_2', \cdots, L_n, L_n', \cdots, L_N, L_N'\}, \quad (13)$$

making each two adjacent single photons $L_n$ and $L_n'$ $(n \in \{1, 2, \cdots, N\})$ in the same state. Here, each single photon is randomly in one of the sixteen states $|\phi\rangle = |\phi\rangle_P \otimes |\phi\rangle_S$, where

$$|\phi\rangle_P \in \{|H\rangle, |V\rangle, |R\rangle, |A\rangle\}, |\phi\rangle_S \in \{|b_1\rangle, |b_2\rangle, |s\rangle, |a\rangle\}. \quad (14)$$

Then, Bob prepares $\delta_1 + \delta_2$ decoy single photons in both polarization and spatial-mode degrees of freedom, each of which is also randomly in one of the above sixteen states, and randomly inserts them into sequence $S$. As a result, a new sequence $S'$ forms. Finally, Bob sends sequence $S'$ to Alice by using the block transmission method [5]. Note that the decoy photon technique [38-39] is employed to check the security of quantum channel here.

**Step 2: The first security check process.** After Alice confirms the receipt of sequence $S'$, they implement the first security check process: (1) Bob tells Alice the positions and the preparation bases of $\delta_1$ decoy single photons (the preparation base of each decoy single photon is one of the four bases $\{Z_P \otimes Z_S, Z_P \otimes X_S, X_P \otimes Z_S, X_P \otimes X_S\}$); (2) Alice uses the bases Bob told to measure $\delta_1$ decoy single photons and informs Bob of her measurement outcomes; (3) Bob determines whether the quantum channel is secure or not by comparing Alice's measurement outcomes with the initial states of $\delta_1$ decoy single photons. If the quantum channel is secure, the communication will be continued; otherwise, it will be terminated.

**Step 3: Alice's encoding and transmission.** Alice discards $\delta_1$ decoy single photons in sequence $S'$. Bob tells Alice the positions of $\delta_2$ decoy single photons. Alice picks out $\delta_2$ decoy single photons, and restores the remaining $2N$ single photons as sequence $S$. Alice divides sequence $S$ into $N$ message single photon groups in the manner that two adjacent single photons $L_n$ and $L_n'$ $(n = 1, 2, \cdots, N)$ form one group. Alice and Bob agree on that only the first single photon in each message single photon group is used for encoding. Then, Alice encodes her two classical bits $(i_n, j_n)$ on the single photon $L_n$ in the $n^{\text{th}}$ message single photon group by performing the composite unitary operation $C_{i_n j_n}$. As a result, the single photon $L_n$ is turned into $C_{i_n j_n} L_n$. Consequently, the $n^{\text{th}}$ message single photon group is changed into $(C_{i_n j_n} L_n, L_n')$. Then, Alice picks out the first single photon from each message single photon group to form a new sequence, i.e.,

$$L = \{C_{i_1 j_1} L_1, C_{i_2 j_2} L_2, \cdots, C_{i_n j_n} L_n, \cdots, C_{i_N j_N} L_N\}. \quad (15)$$

The remaining single photon from each message single photon group forms another new sequence, i.e.,

$$L' = \{L_1', L_2', \cdots, L_n', \cdots, L_N'\}. \quad (16)$$

In order to encode her checking bits, Alice randomly performs one of the four composite unitary operations of formula (12) on each of $\delta_2$ decoy single photons. Afterward, Alice randomly inserts these encoded decoy single photons into sequence $L$ to form a new sequence $L''$. Finally, Alice transmits sequence $L''$ to Bob also in the manner of block transmission [5], and keeps $L'$ in her hand.

**Step 4: The second security check process.** After Bob confirms Alice the receipt of sequence $L''$, they implement the second security check process: (1) Alice tells Bob the positions of the $\delta_2$ encoded decoy single photons in sequence $L''$; (2) Since he generates the $\delta_2$ decoy single photons by himself, Bob can know their initial states and measuring bases. Bob uses the right measuring bases to measure them and decodes out Alice's checking bits. Then, Bob informs Alice of his decoding outcomes; (3) Alice compares her checking bits with Bob's decoding outcomes. If there is no error, the communication will be continued; otherwise, it will be terminated.

**Step 5: Bob's encoding and their decoding.** Bob drops out the $\delta_2$ encoded decoy single photons in sequence $L''$ to restore sequence $L$. Then, Bob encodes his two classical bits $(k_n, l_n)$ on the single photon $C_{i_n j_n} L_n$ in the $n^{\text{th}}$ message single photon group by performing the composite unitary operation $C_{k_n l_n}$. As a result, the single photon $C_{i_n j_n} L_n$ is turned into a new single photon $C_{k_n l_n} C_{i_n j_n} L_n$. Accordingly, sequence $L$ is changed into

$$L = \{C_{k_1 l_1} C_{i_1 j_1} L_1, C_{k_2 l_2} C_{i_2 j_2} L_2, \cdots, C_{k_n l_n} C_{i_n j_n} L_n, \cdots, C_{k_N l_N} C_{i_N j_N} L_N\}. \quad (17)$$

Since Bob generates the single photon $L_n$ by himself, he naturally knows its initial state and the measuring base of single photon $C_{k_n l_n} C_{i_n j_n} L_n$. Bob uses the correct measuring base to measure the single photon $C_{k_n l_n} C_{i_n j_n} L_n$. Afterward, Bob publishes his measurement outcome of the single photon $C_{k_n l_n} C_{i_n j_n} L_n$, whose announcement needs four classical bits. With the initial state of the single photon $L_n$ and his own composite unitary operation $C_{k_n l_n}$, Bob can decode out Alice's two classical bits $(i_n, j_n)$. As for Alice, according to Bob's announcement on the measurement outcome of the single photon $C_{k_n l_n} C_{i_n j_n} L_n$, she can choose the correct measuring base to measure the single photon $L'_n$ in sequence $L'$. As a result, she knows the initial state of the single photon $L_n$, since each two adjacent single photons $L_n$ and $L'_n$ is generated in the same state by Bob. With her own composite unitary operation $C_{i_n j_n}$, Alice can also decode out Bob's two classical bits $(k_n, l_n)$.

Now a concrete example is given to explain the bidirectional communication process of the above protocol, taking the $n$ th group $(L_n, L'_n)$ as an example. Suppose that $(i_n, j_n) = (0,0)$ and $(k_n, l_n) = (0,1)$. Moreover, assume that $L_n$ and $L'_n$ are initially prepared in the state of $|H\rangle \otimes |s\rangle$. As a result, after Alice and Bob's encoding, $L_n$ is turned into

$$C_{k_n l_n} C_{i_n j_n} L_n = (I_P \otimes U_S)(I_P \otimes I_S)(|H\rangle \otimes |s\rangle) = -|H\rangle \otimes |a\rangle, \quad (18)$$

while $L'_n$ is kept stationary. Since Bob generates $L_n$ by himself, he naturally knows its initial state and the measuring base of $C_{k_n l_n} C_{i_n j_n} L_n$. Afterward, Bob uses the right measuring base $Z_P \otimes X_S$ to measure $C_{k_n l_n} C_{i_n j_n} L_n$, and publishes his measurement outcome. According to the initial state of $L_n$ and his own composite unitary operation $C_{k_n l_n}$, Bob can decode out that $(i_n, j_n) = (0,0)$. As for Alice, after hearing of Bob's publishment on the measurement outcome of $C_{k_n l_n} C_{i_n j_n} L_n$, she uses the right measuring base $Z_P \otimes X_S$ to measure $L'_n$ and knows the initial state of $L_n$. Then, Alice can decode out that $(k_n, l_n) = (0,1)$ via her own composite unitary operation $C_{i_n j_n}$.

## 3 Security analysis

(1) Analysis on the information leakage problem

Without loss of generality, we still use the above example to analyze the information leakage problem here. The relations among Bob's measurement outcome of $C_{k_n l_n} C_{i_n j_n} L_n$, Alice's composite unitary operation $C_{i_n j_n}$ and Bob's composite unitary operation $C_{k_n l_n}$ are summarized in Tables 1-4 when the composite preparing base of $L_n$ is $Z_P \otimes X_S$. In each table, the first row denotes Alice's composite unitary operation $C_{i_n j_n}$ while the first column is Bob's composite unitary operation $C_{k_n l_n}$. After Bob's publishment, Eve is aware of the state of $C_{k_n l_n} C_{i_n j_n} L_n$ and the composite preparing base of $L_n$, but still has no knowledge about the initial state of $L_n$. As a result, she has to randomly guess the initial state of $L_n$. If she guesses that the initial state of $L_n$ is $|H\rangle \otimes |s\rangle$, according to Table 1, Eve will think that $\{(i_n, j_n), (k_n, l_n)\}$ are one of

$$\{(0,0),(0,1)\}, \{(0,1),(0,0)\}, \{(1,0),(1,1)\}, \{(1,1),(1,0)\}; \quad (19)$$

if she guesses that the initial state of $L_n$ is $|H\rangle \otimes |a\rangle$, according to Table 2, she will think that $\{(i_n, j_n), (k_n, l_n)\}$ are one of

$$\{(0,0),(0,0)\}, \{(0,1),(0,1)\}, \{(1,0),(1,0)\}, \{(1,1),(1,1)\}; \quad (20)$$

if she guesses that the initial state of $L_n$ is $|V\rangle \otimes |s\rangle$, according to Table 3, she will think that $\{(i_n, j_n), (k_n, l_n)\}$ are one of

$$\{(0,0),(1,1)\}, \{(0,1),(1,0)\}, \{(1,0),(0,1)\}, \{(1,1),(0,0)\}; \quad (21)$$

if she guesses that the initial state of $L_n$ is $|V\rangle \otimes |a\rangle$, according to Table 4, she will think that $\{(i_n, j_n), (k_n, l_n)\}$ are one of

$$\{(0,0),(1,0)\}, \{(0,1),(1,1)\}, \{(1,0),(0,0)\}, \{(1,1),(0,1)\}. \quad (22)$$

As a result, there are totally sixteen kinds of uncertainty for Eve, which includes

$$-\sum_{i=1}^{16} p_i \log_2 p_i = -16 \times \frac{1}{16} \log_2 \frac{1}{16} = 4 \qquad (23)$$

bit information, from the viewpoint of Shannon's information theory [40]. This amount of information is equal to that of Alice and Bob's secret bits. Thus, no information is leaked out to Eve. It is easy to know that, during the communication process, $L_n'$ plays the role of an auxiliary single photon for privately sharing the initial state of $L_n$ between Alice and Bob so that Eve is unable to know the initial state of $L_n$. In this way, although Eve is aware of the state of $C_{k_n l_n} C_{i_n j_n} L_n$ from Bob's publishment, she still knows nothing about Alice and Bob's secret bits.

Table 1  The relations among Bob's measurement outcome of $C_{k_n l_n} C_{i_n j_n} L_n$, Alice's composite unitary operation $C_{i_n j_n}$ and Bob's composite unitary operation $C_{k_n l_n}$ when the initial state of $L_n$ is $|H\rangle \otimes |s\rangle$

|  | $I_P \otimes I_S$ | $I_P \otimes U_S$ | $U_P \otimes I_S$ | $U_P \otimes U_S$ |
|---|---|---|---|---|
| $I_P \otimes I_S$ | $|H\rangle \otimes |s\rangle$ | $|H\rangle \otimes |a\rangle$ | $|V\rangle \otimes |s\rangle$ | $|V\rangle \otimes |a\rangle$ |
| $I_P \otimes U_S$ | $|H\rangle \otimes |a\rangle$ | $|H\rangle \otimes |s\rangle$ | $|V\rangle \otimes |a\rangle$ | $|V\rangle \otimes |s\rangle$ |
| $U_P \otimes I_S$ | $|V\rangle \otimes |s\rangle$ | $|V\rangle \otimes |a\rangle$ | $|H\rangle \otimes |s\rangle$ | $|H\rangle \otimes |a\rangle$ |
| $U_P \otimes U_S$ | $|V\rangle \otimes |a\rangle$ | $|V\rangle \otimes |s\rangle$ | $|H\rangle \otimes |a\rangle$ | $|H\rangle \otimes |s\rangle$ |

Table 2  The relations among Bob's measurement outcome of $C_{k_n l_n} C_{i_n j_n} L_n$, Alice's composite unitary operation $C_{i_n j_n}$ and Bob's composite unitary operation $C_{k_n l_n}$ when the initial state of $L_n$ is $|H\rangle \otimes |a\rangle$

|  | $I_P \otimes I_S$ | $I_P \otimes U_S$ | $U_P \otimes I_S$ | $U_P \otimes U_S$ |
|---|---|---|---|---|
| $I_P \otimes I_S$ | $|H\rangle \otimes |a\rangle$ | $|H\rangle \otimes |s\rangle$ | $|V\rangle \otimes |a\rangle$ | $|V\rangle \otimes |s\rangle$ |
| $I_P \otimes U_S$ | $|H\rangle \otimes |s\rangle$ | $|H\rangle \otimes |a\rangle$ | $|V\rangle \otimes |s\rangle$ | $|V\rangle \otimes |a\rangle$ |
| $U_P \otimes I_S$ | $|V\rangle \otimes |a\rangle$ | $|V\rangle \otimes |s\rangle$ | $|H\rangle \otimes |a\rangle$ | $|H\rangle \otimes |s\rangle$ |
| $U_P \otimes U_S$ | $|V\rangle \otimes |s\rangle$ | $|V\rangle \otimes |a\rangle$ | $|H\rangle \otimes |s\rangle$ | $|H\rangle \otimes |a\rangle$ |

Table 3  The relations among Bob's measurement outcome of $C_{k_n l_n} C_{i_n j_n} L_n$, Alice's composite unitary operation $C_{i_n j_n}$ and Bob's composite unitary operation $C_{k_n l_n}$ when the initial state of $L_n$ is $|V\rangle \otimes |s\rangle$

|  | $I_P \otimes I_S$ | $I_P \otimes U_S$ | $U_P \otimes I_S$ | $U_P \otimes U_S$ |
|---|---|---|---|---|
| $I_P \otimes I_S$ | $|V\rangle \otimes |s\rangle$ | $|V\rangle \otimes |a\rangle$ | $|H\rangle \otimes |s\rangle$ | $|H\rangle \otimes |a\rangle$ |
| $I_P \otimes U_S$ | $|V\rangle \otimes |a\rangle$ | $|V\rangle \otimes |s\rangle$ | $|H\rangle \otimes |a\rangle$ | $|H\rangle \otimes |s\rangle$ |
| $U_P \otimes I_S$ | $|H\rangle \otimes |s\rangle$ | $|H\rangle \otimes |a\rangle$ | $|V\rangle \otimes |s\rangle$ | $|V\rangle \otimes |a\rangle$ |
| $U_P \otimes U_S$ | $|H\rangle \otimes |a\rangle$ | $|H\rangle \otimes |s\rangle$ | $|V\rangle \otimes |a\rangle$ | $|V\rangle \otimes |s\rangle$ |

Table 4  The relations among Bob's measurement outcome of $C_{k_n l_n} C_{i_n j_n} L_n$, Alice's composite unitary operation $C_{i_n j_n}$ and Bob's composite unitary operation $C_{k_n l_n}$ when the initial state of $L_n$ is $|V\rangle \otimes |a\rangle$

|  | $I_P \otimes I_S$ | $I_P \otimes U_S$ | $U_P \otimes I_S$ | $U_P \otimes U_S$ |
|---|---|---|---|---|
| $I_P \otimes I_S$ | $|V\rangle \otimes |a\rangle$ | $|V\rangle \otimes |s\rangle$ | $|H\rangle \otimes |a\rangle$ | $|H\rangle \otimes |s\rangle$ |
| $I_P \otimes U_S$ | $|V\rangle \otimes |s\rangle$ | $|V\rangle \otimes |a\rangle$ | $|H\rangle \otimes |s\rangle$ | $|H\rangle \otimes |a\rangle$ |
| $U_P \otimes I_S$ | $|H\rangle \otimes |a\rangle$ | $|H\rangle \otimes |s\rangle$ | $|V\rangle \otimes |a\rangle$ | $|V\rangle \otimes |s\rangle$ |
| $U_P \otimes U_S$ | $|H\rangle \otimes |s\rangle$ | $|H\rangle \otimes |a\rangle$ | $|V\rangle \otimes |s\rangle$ | $|V\rangle \otimes |a\rangle$ |

(2) Analysis of Eve's active attacks

During the whole communication, the single photon $L_n$ is transmitted forth and back, thus twice security check processes are conducted totally. Obviously, the second security check process uses the message authentication method to check the existence of Eve during the transmission of sequence $L^{"}$ from Alice to Bob. Because she has no knowledge about the positions and the initial states of the single photons in sequence $L^{"}$, even if she intercepts sequence $L^{"}$, Eve can not obtain anything useful about Alice's bits but disturb its transmission. As a result, Eve's attack behavior can be detected by the second security check process inevitably. Thus, the security of the proposed QD protocol is determined by the first security check process, which adopts the decoy photon technique [38-39] to detect the attack behavior of Eve. It is well known that the decoy photon technique [38-39] can be thought as a variation of the security check method of the BB84 QKD protocol [1], which has been proven to be unconditionally secure [41]. Now we validate its effectiveness against Eve's several active attacks as follows.

① The Trojan horse attacks

There are two kinds of Trojan horse attack strategies, i.e., the invisible photon eavesdropping [42] and the delay-photon Trojan horse attack [43-44]. In order to overcome the invisible photon eavesdropping, when she receives sequence $S^{'}$ from Bob, Alice can insert a filter in front of her devices to filter out the photon signal with an illegitimate wavelength [44-45]; in order to resist the delay-photon Trojan horse attack, Alice can adopt a photon number splitter (PNS:50/50) to split each sampling signal of $\delta_1$ decoy single photons and use proper measuring bases to measure the two signals after the PNS [44-45]. If the multiphoton rate is abnormally high, the communication will be terminated; otherwise, the communication will be continued.

② The intercept-resend attack

Eve prepares a fake sequence beforehand, which is composed of the single photons randomly in one of the sixteen states $|\phi\rangle = |\phi\rangle_P \otimes |\phi\rangle_S$, where

$$|\phi\rangle_P \in \{|H\rangle, |V\rangle, |R\rangle, |A\rangle\}, |\phi\rangle_S \in \{|b_1\rangle, |b_2\rangle, |s\rangle, |a\rangle\}. \tag{24}$$

After she intercepts sequence $S^{'}$, Eve uses her fake sequence to replace it and resends the new sequence to Alice. However, before Bob tells Alice the positions and the preparation bases of $\delta_1$ decoy single photons, Eve has no knowledge about them. As a result, Eve's attack can be detected by the first security check process with the probability of

$$1 - \left(\frac{1}{4}\right)^{\delta_1}, \tag{25}$$

as Alice's measurement outcomes on the fake decoy single photons are not always identical with the genuine ones.

③ The measure-resend attack

After she intercepts sequence $S^{'}$, Eve measures each of its single photons randomly with one of the four bases $\{Z_P \otimes Z_S, Z_P \otimes X_S, X_P \otimes Z_S, X_P \otimes X_S\}$ and resends the new sequence to Alice. Because Eve's measuring bases for decoy single photons are not always consistent with Bob's preparing bases for them, her attack can be detected by the first security check process with the probability of

$$1 - \left(\frac{9}{16}\right)^{\delta_1}. \tag{26}$$

④ The entangle-measure attack

Eve may steal partial information by entangling her auxiliary particle $|\varepsilon_i\rangle$ with the particle in sequence $S^{'}$ through a unitary operation $\hat{U}_E$. Without loss of generality, we take one decoy single photon in the state of $|H\rangle \otimes |b_1\rangle$ for example to analyze the entangle-measure attack. It follows that

$$\hat{U}_E |\varepsilon_i\rangle_E \otimes (|H\rangle \otimes |b_1\rangle) = \alpha |\varepsilon_i\rangle_E |Hb_1\rangle + \beta |\varepsilon_{i\oplus 1}\rangle_E |Vb_1\rangle, \tag{27}$$

where

$$|\alpha|^2 + |\beta|^2 = 1, \tag{28}$$

$|\varepsilon_i\rangle_E$ and $|\varepsilon_{i\oplus 1}\rangle_E$ are pure ancilla states uniquely determined by $\hat{U}_E$ and

$$_E\langle\varepsilon_i|\varepsilon_{i\oplus 1}\rangle_E = 0. \tag{29}$$

As a result, when Alice performs the first security check process by measuring this state with the base of $Z_P \otimes Z_S$, Eve can be detected with the probability of $|\beta|^2$.

To sum up, it can be concluded now that the proposed QD protocol can overcome Eve's active attacks.

## 4 Discussion and conclusion

(1) The information-theoretical efficiency

The information-theoretical efficiency defined by Cabello [3] is

$$\eta = \frac{v_c}{q_t + v_t}, \qquad (30)$$

where $v_c$, $q_t$ and $v_t$ are the expected secret bits received, the qubits used and the classical bits exchanged between two communicants, respectively. In the proposed QD protocol, after ignoring the two security check processes, two adjacent single photons $L_n$ and $L_n^{'}$ can be used for exchanging two classical bits $(i_n, j_n)$ from Alice and two classical bits $(k_n, l_n)$ from Bob, while four classical bits are consumed by Bob's announcement on his measurement outcome of the single photon $C_{k_n l_n} C_{i_n j_n} L_n$. As a result, it follows that $v_c = 4$, $q_t = 4$, $v_t = 4$, making $\eta = \frac{4}{4+4} \times 100\% = 50\%$.

(2) Comparisons of two previous QD protocols

Recently, Wang et al. [35] also constructed a QD protocol with single photons in both polarization and spatial-mode degrees of freedom. However, later, Zhang and Situ [36] pointed out that the QD protocol of Ref.[35] runs the risk of information leakage, and remedied this drawback by modifying Alice and Bob's encoding rules. In Zhang and Situ's improved protocol [36], after ignoring the security check processes, one single photon in both polarization and spatial-mode degrees of freedom can be used for exchanging one classical bit from Alice and one classical bit from Bob, while two classical bits are consumed by Alice's announcement on the value of $R$. As a result, it follows that $v_c = 2$, $q_t = 2$, $v_t = 2$, making $\eta = \frac{2}{2+2} \times 100\% = 50\%$. Therefore, the proposed QD protocol has the same information-theoretical efficiency to Zhang and Situ's improved protocol [36]. However, Zhang and Situ's improved protocol [36] adopts seven different composite unitary operations, while the proposed protocol only needs four. In addition, the proposed protocol provides a new method to overcome the information leakage problem of QD protocol working with single photons in both polarization and spatial-mode degrees of freedom, i.e, the method of directly transmitting auxiliary single photons in both polarization and spatial-mode degrees of freedom from one communicant to another.

In addition, Shi et al. [29] put forward an information leakage resistant QD protocol based on single photons in only one degree of freedom. In Shi et al.'s QD protocol [29], after ignoring the security check processes, two single photons in only one degree of freedom can be used for exchanging one classical bit from Alice and one classical bit from Bob, while two classical bits are consumed by Bob's announcement on his measurement outcome. As a result, it follows that $v_c = 2$, $q_t = 2$, $v_t = 2$, making $\eta = \frac{2}{2+2} \times 100\% = 50\%$. Therefore, the proposed QD protocol has the same information-theoretical efficiency to Shi et al.'s QD protocol [29]. However, two single photons in only one degree of freedom only carry two classical bits in total in Shi et al.'s QD protocol [29], while two single photons in both polarization and spatial-mode degrees of freedom can totally carry four classical bits in the proposed QD protocol. Therefore, the proposed QD protocol doubles the capacity of quantum communication of Shi et al.'s QD protocol [29].

To sum up, we design a novel QD protocol with single photons in both polarization and spatial-mode degrees of freedom. The proposed QD protocol overcomes the information leakage problem by directly transmitting the auxiliary single photons in both polarization and spatial-mode degrees of freedom from one communicant to another, which makes two communicants privately share the initial states of their counterparts used for encoding. We also validate in detail that the proposed QD protocol can overcome the Trojan horse attack, the intercept-resend attack, the measure-resend attack and the entangle-measure attack from Eve. The proposed QD protocol only adopts single photons in both polarization and spatial-mode degrees of freedom as quantum resource and single-photon measurements. Thus, it has good feasibility in practice, as the preparation and the quantum measurement of a single photon in both polarization and spatial-mode degrees of freedom can be realized with current experimental techniques. Compared with Zhang and Situ's improved protocol [36], the proposed QD protocol decreases the number of composite unitary operations used for encoding; and compared with Shi et al.'s QD protocol [29], the proposed QD protocol doubles the capacity of quantum communication. The influence of noise on single photon in both polarization and spatial-mode degrees of freedom is very complicated, so we leave this problem for further research.

**Acknowledgements**

The authors would like to thank the anonymous reviewers for their valuable comments that help enhancing the quality of this paper. Funding by the National Natural Science Foundation of China (Grant No.62071430) and Zhejiang Gongshang University, Zhejiang Provincial Key Laboratory of New Network Standards and Technologies (No. 2013E10012) is gratefully acknowledged.